# DETERMINATION OF CHARACTERISTICS OF NEWLY DISCOVERED ECLIPSING BINARY 2MASS J18024395 +4003309 = VSX J180243.9+400331

I.L. Andronov[1], V.V. Breus[1], S. Zoła[2,3]

[1] Department "High and Applied Mathematics", Odessa National Maritime University, Odessa, Ukraine, *tt_ari @ukr.net, bvv_2004@ua.fm*
[2] Astronomical Observatory, Jagiellonian University, Cracow, Poland
[3] Mt. Suhora Observatory, Pedagogical University, Cracow, Poland

ABSTRACT. During processing the observations of the intermediate polar 1RXS J180340.0+401214, obtained 26.05.2012 at the 60-cm telescope of the Mt. Suhora observatory (Krakow, Poland), variability of 2MASS J18024395 +4003309 was discovered. As this object was not listed in the "General Catalogue of Variale Stars" or "Variable Stars Index", we registered it as VSX J180243.9+400331. Additionally we used 189 separate observations from the Catalina Sky Survey spread over 7 years. The periodogram analysis yields the period of $0^d.3348837 \pm 0^d.0000002$.

The object was classified as the Algol-type eclipsing binary with a strong effect of ellipticity. The depths of the primary and secondary minima are nearly identical, which corresponds to a brightness (and maybe) mass ratio close to 1. The statistically optimal degree of the trigonometric polynomial $n=4$. The most recent minimum occurred at HJD 2456074.4904. The brightness range from our data is 16.56-17.52 (V), 16.18-17.08 (R).

The NAV ("New Algol Variable") algorithm was applied for statistically optimal phenomenological modeling and determination of corresponding parameters.

During our monitoring of the intermediate polar 1RXS J180340.0+401214 (=RXJ1803), Breus (2012) discovered a new variable, which was absent in the "General Catalogue of Variable Stars" (Samus' et al. 2012) and "Variable Stars Index" (2012). The CCD images were taken by S.Zoła using V and R filters at the 60-cm telescope of the Mt.Suhora Observatory. Contrary to previous studies of the same field (cf. Andronov et al. 2011) with smaller field, these images contained this new variable. The variable was registered as VSX J180243.9+400331 (hereafter VSX1802). In this field, the brightness of some stars was calibrated by Henden (2005). The brightness of the "main" comparison star "1" is V= $14.807^m$, Rc=$14.436^m$. For better accuracy, we have used the "multiple comparison stars" method (Andronov and Baklanov 2004, Kim et al., 2004). The color transformation equations are:
$V_I-V_H$=0.031-0.061(V-Rc),   $R_I-R_H$=-0.002-0.101(V-Rc), $(V-R)_I$=0.035+0.855(V-Rc), $(V-Rc)$=-0.028+1.144$(V-R)_I$

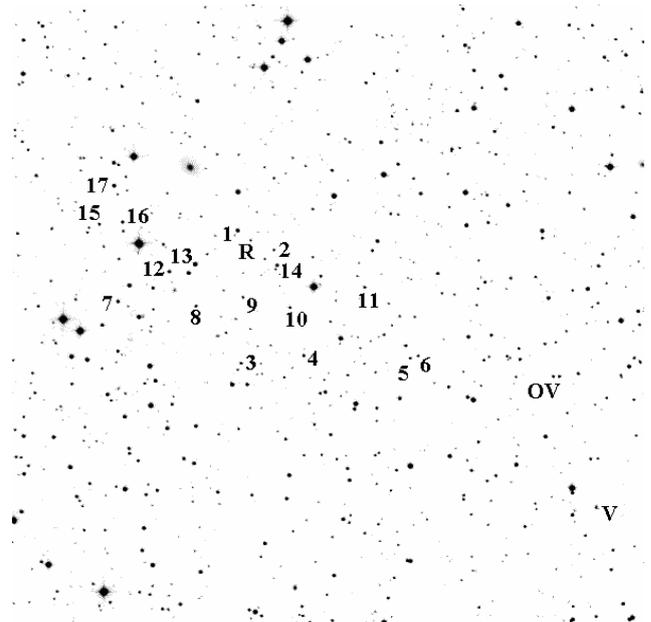

Fig. 1. Finding chart (20') with multiple comparison stars. The new variable is marked as "V" and RXJ1803 as "R", and OV Her as "OV". North is up and West is left. The numeration of comparison stars is as in Andronov et al. (2011), except the star "13".

with corresponding error estimates 0.013, 0.025, 0.020, 0.039, 0.018, 0.034, 0.021, 0.045. For the color index, the deviation of the coefficient from unity exceeds 4s and thus is to be taken into account. We recommend the same stars to be used at other telescopes for further joint analysis of the data.

Additionally we have found published data from the Catalina survey (Drake et al. 2009). Although they are noisy, the best accuracy for the period was obtained using the set "Catalina"+(scaled)"Suhora R". The parameters of the best trigonometric polynomial fit (Andronov, 1994, 2003) are listed in the abstract.

Although the star is not listed in the GCVS and VSX, it was independently found by Parimucha et al. (2011) and named "Kol5". They published brightness $Max_I$=16.13, $Max_{II}$=16.16, $min_I$=16.89, $min_{II}$=16.80 However, their period estimate was wrong, as one may see from the phase light curves. The minima are nearly of the same depth, and their "primary" minimum was in fact a secondary one leading to a 0.5 cycle miscount during a year. With a corrected ephemeris, all minima occur at proper positions.

A funny coincidence, but the cycle miscount during one year also was found for the "main" star RXJ1803, for which Breus et. al. (2012) published corrected elements

$T_{max}$= 2454604.04449(14) +0.017596986(3)$E$

for the spin maxima.

The phase curve in VR is only partial, and only a primary minimum was covered. One should note an apparently larger amplitude of these recent observations as compared with that at the Catalina survey, thus we had to scale R data for further periodogram analysis.

To smooth the complete phase curve using the Catalina data only, we used the NAV algorithm (Andronov 2012). The best fit values are $C_8$=D/2=0.119 (filter half-width), $\beta_1$=$\beta_2$=1, mean brightness out of eclipse $C_1$=16.236$^m$(5), semi-amplitude of ellipsoidal variations $C_3$=0.078$^m$(7), unbiased depths of the primary ($C_6$=0.568$^m$(21)) and secondary ($C_7$=0.486$^m$(23)) minima. Coefficients $C_2$, $C_4$, $C_5$, which describe a reflection and O'Connell effect, are not statistically significant. The corresponding light curve is shown in Fig. 3.

The parameter Y (Andronov 2012), which is related to degree of eclipses (0 – "no eclipses", 1 – "both eclipses are full"), is equal to 0.768(18). This value is closer to "full eclipses" rather than "no eclipses". Another parameter corresponds to an effective ratio of the surface brightnesses $i_1/i_2$=1.130(53), i.e. exceeds unity by ~2σ only. However, the difference in brightness of minima is $min_I$-$min_{II}$=0.083$^m$(26) is more statistically significant.

Corresponding brightness from the fit $Max_I$=16.155(6), $Max_{II}$=16.160(6), $min_I$=16.882(12), $min_{II}$=16.813(11) are close to that determined by Parimucha et al. (2011), but the accuracy estimate is better by a factor of ~5.

*Acknowlwdgements.* We thank Dr. L.Hric for helpful discussions and hospitality during the stellar conference Bezovec-2012 and to the Queen Jadwiga Foundation for individual grants to I.L.A. and B.V.V. The study was made in a course of the projects "Inter-Longitude Astronomy" (Andronov et al., 2010) and "Ukrainian Virtual Observatory" (Vavilova et al., 2012).

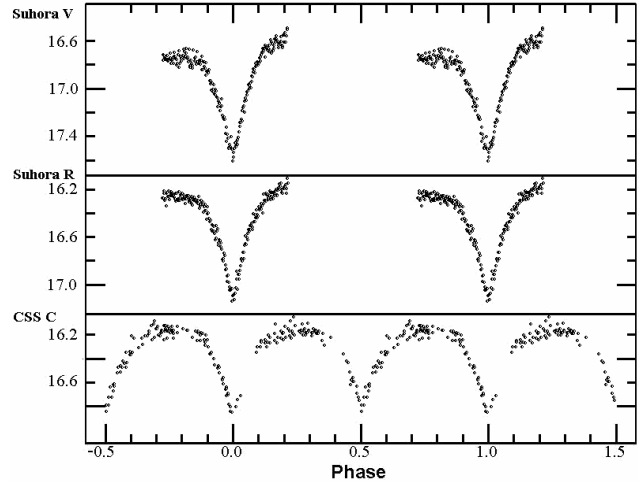

Fig. 2. Phase light curves of VSX1802

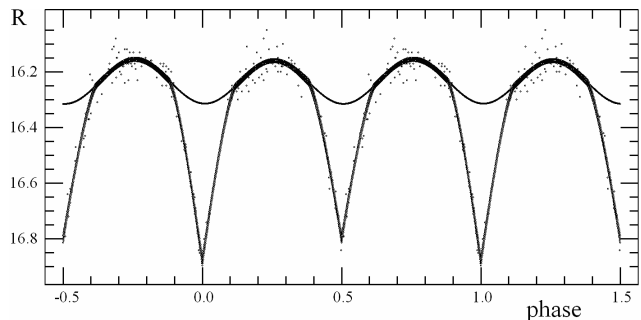

Fig. 3. The "NAV" fit for the phase curve and a corresponding "1σ" corridor. At the phases of eclipses, an additional curve corresponding to a continuation of "out of eclipse" parts is also shown.